\documentclass{rmaa}

\newcommand{\kms}{\mbox{$\,$km s$^{-1}$}}

\title{H91$\alpha$ Radio Recombination Line and 3.5 cm Continuum
Observations of the Planetary Nebula NGC~3242}

 \author{Luis F. Rodr\'\i guez and Yolanda G\'omez
  \affil{Centro de Radioastronom\'{\i}a y Astrof\'{\i}sica, UNAM, Morelia}
\and J. Alberto L\'opez, Ma. Teresa Garc\'\i a-D\'\i az and David M. Clark
\affil{Instituto de Astronom\'{\i}a, UNAM, Ensenada}
 }

\fulladdresses{
\item Yolanda G\'omez and Luis F. Rodr\'\i guez: Centro de Radiostronom\'{\i}a
y Astrof\'\i sica, 
UNAM, A. P. 3-72, 
(Xangari), 58089 Morelia, Michoac\'an, M\'exico
(y.gomez; l.rodriguez@crya.unam.mx).
\item David M. Clark, Ma. Teresa Garc\'\i a-D\'\i az, and J. Alberto L\'opez: 
Instituto de Astronom\'\i a, Universidad Nacional Aut\'onoma de M\'exico 
Campus Ensenada, Ensenada, Baja California, 22800, M\'exico
(jal; tere; dmclark@astrosen.unam.mx).
}

\shortauthor{Rodr\'\i guez et al.}
\shorttitle{Radio Recombination Line Emission from NGC 3242}

\SetVolume{00} \SetFirstPage{001} \SetYear{2010}
\ReceivedDate{\today} 
\AcceptedDate{Year Month Day} 

\resumen{Presentamos observaciones de 
alta sensitividad de la l\'\i nea de recombinaci\'on de radio
H91$\alpha$ a 3.5 cm y de su continuo adyacente hacia la nebulosa planetaria
NGC~3242. La temperatura electr\'onica derivada suponiendo
equilibrio termodin\'amico local es consistente dentro de $\sim$10\%
con la determinada
de l\'\i neas \'opticas y de la discontinuidad de Balmer.
La emisi\'on de l\'\i nea y la de continuo tienen una distribuci\'on
espacial muy similar, indicando que a esta longitud de onda no hay otro proceso de emisi\'on
de continuo presente de manera significativa.
En particular concluimos que la emisi\'on de polvo rotante no es importante 
a esta longitud de onda.
En esta l\'\i nea de recombinaci\'on de radio la nebulosa presenta una estructura en
velocidad radial consistente con la obtenida de observaciones de l\'\i neas en el visible.
}
 
\abstract{We present high sensitivity
H91$\alpha$ and 3.5 cm radio continuum observations toward
the planetary nebula NGC~3242.
The electron temperature determined assuming local thermodynamic equilibrium
is consistent within $\sim$10\%
with that derived from optical lines and the Balmer discontinuity.
The line emission and the continuum emission have very similar spatial
distribution, suggesting that at this wavelength there is no other continuum process
present in a significant manner.
In particular, we conclude that emission from spinning dust is not important
at this wavelength.
In this radio recombination line the nebula presents a radial velocity structure consistent
with that obtained from observations of optical lines.
}
      

\begin{document}

\maketitle

\section{Introduction}

NGC 3242 (PN G261.0+32.0)
is a multiple-shell, attached-halo planetary nebula (PN), located 
at a distance of about 1 kpc 
(Stanghellini \& Pasquali 1995). Recently, an expansion distance measured with radio continuum
observations (Hajian, Phillips, \& Terzian 1995) placed this PN at 0.42$\pm$0.16 kpc.
As it is commonly the case in planetary nebula, the individual distances
show a large spread, ranging from 0.28 to 2.0 kpc for NGC~3242 (Acker et al. 1992).
We will adopt the distance of 1 kpc proposed by Stanghellini \& Pasquali (1995).
The inner and outer shells of the object have diameters of 
15{''} and 29{''}, respectively. NGC 3242 is a high-excitation planetary nebula (class 7; 
Pottasch 1984) powered by a $\geq$1000 $L_\odot$
(Stanghellini \& Pasquali 1995) SD0 star (Heap 1986) with 
a surface temperature of $T_{eff}$ = 60,000 K 
(Acker et al. 1992).
The post-AGB mass of the central star is estimated to be $M_{cs} = 0.56 \pm 0.01~ M_\odot$, 
corresponding to a main-sequence mass of $M_{ms} = 1.2 \pm 0.2 ~ M_\odot$ (Galli et al. 1997). 

NGC 3242 was the first PN with a firm detection of the $^3He^+$ hyperfine line at 8.665 GHz
(Balser, Rood, \& Bania 1999). As part of these $^3He^+$ studies, the Very Large Array archive
contains also high quality, unpublished data of the H91$\alpha$ radio recombination
line (project AR271), that has a nearby
rest frequency of 8.584 GHz. In this paper we analyze this radio recombination
line and the adjacent continuum. The motivation of this analysis is in first place to
study the radio recombination line emission of this planetary nebula at interferometer
angular resolutions since to our knowledge only single dish observations have been 
published. The second reason to study the continuum and radio recombination line emission
of this PN is that recently (Casassus 2005; 2007) it 
was proposed as one of the planetary nebulae that may
show excess anomalous microwave emission (e. g. Casassus et al. 2004; Finkbeiner et al. 2002).
This new continuum radiation mechanism has been 
attributed to electric dipole emission from rapidly 
rotating dust grains ("spinning dust"), as predicted by Draine \& Lazarian (1998). 
In a nebula that is optically thin at centimeter wavelengths one expects the free-free
continuum emission and the radio recombination line emission to be very similar
since both originate mostly from electron-proton interactions. The significant
presence of an independent
radiation mechanism such as spinning dust emission is expected to become evident in a
careful comparison between the total continuum and the radio recombination line emissions.

\section{Observations}

The observations were made at 3.5 cm during 1992 August 15 in the D
configuration
of the VLA of the NRAO\footnote{The National Radio 
Astronomy Observatory is operated by Associated Universities 
Inc. under cooperative agreement with the National Science Foundation.}.
The frequency of the observations was centered at
the rest frequency of the H91$\alpha$ radio recombination line,
8.584 GHz.
The source 1331+305 was used as an absolute amplitude
calibrator (with an adopted flux density of 5.10 Jy).
The source 1035-201 was used as the phase calibrator
(with a bootstrapped flux density of 0.563$\pm$0.007 Jy)
and the source 0542+498 was used as the bandpass calibrator
(with a bootstrapped flux density of 4.370$\pm$0.039 Jy).
The observations were made with 31 spectral line channels with a
frequency resolution of 195.3 kHz (corresponding to 6.8 km s$^{-1}$).
A continuum channel (the channel 0) contains the average of the central 75\% of 
the available band. The data were calibrated following the standard VLA
procedures for spectral line, using the software package AIPS of NRAO.
Using the channel 0, the data were self-calibrated in amplitude and phase.
The images were made with natural weighting to obtain the highest
sensitivity. The synthesized beam has half power full width dimensions of
$13\rlap'{''}0 \times 8\rlap.{''}4$, 
with the major axis at a position angle of $+10^\circ$. 

\section{Interpretation and Results}

\subsection{Average LTE Electron Temperature}

In Figure 1 we show the total H91$\alpha$ line emission, integrated
on a box of solid angle $53{''} \times 61{''}$ ($\Delta \alpha \times \Delta \delta$)
that includes all the bright emission from the ionized nebula.
The integrated parameters for line and continuum in this box are given in Table 1.
The continuum parameters were derived from an image made using 14 channels in the
spectrum that were considered to be free of line emission.

\begin{figure}
\centering
\includegraphics[scale=0.4, angle=0]{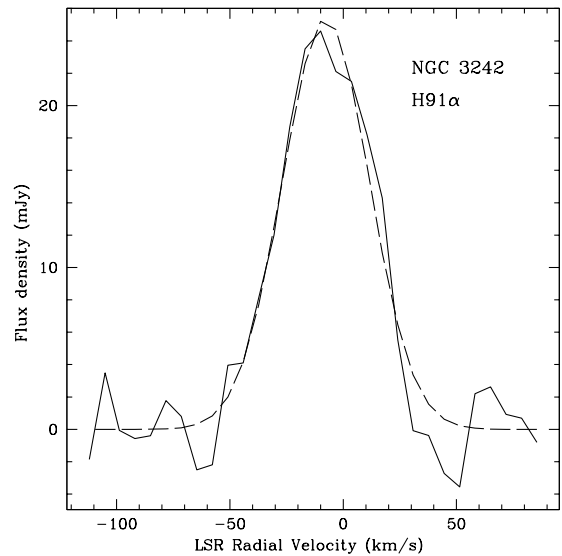}
 \caption{Integrated H91$\alpha$ line emission from the planetary
nebula NGC 3242. The dashed line is the least-squares fitted
Gaussian profile. The parameters of the fit are given in Table 1.}
  \label{fig1}
\end{figure}

\begin{table}[htbp]
  \setlength{\tabnotewidth}{0.9\columnwidth} 
  \tablecols{2} 
  \caption{Continuum and H91$\alpha$ line parameters for NGC~3242}
  \begin{center}
    \begin{tabular}{lc}\hline\hline
Parameter$^a$ &  Value \\
\hline
$S_C$ &  660$\pm$10 mJy \\
$S_L$ &  25.4$\pm$1.1 mJy \\
$\Delta v$ & 45.2$\pm$2.3 km s$^{-1}$ \\
$v_{LSR}$ &  -7.6$\pm$1.0 km s$^{-1}$ \\
\hline\hline
\tabnotetext{a}{$S_C$ = Continuum Flux Density, $S_L$ = Peak Line Flux Density,
$\Delta v$ = Half Maximum Full Width, $v_{LSR}$ = LSR Radial Velocity.}

    \label{tab:1}
    \end{tabular}
  \end{center}
\end{table}

The electron temperature for an optically thin plasma, assuming LTE, is given by:

$$T^*_e = \biggl[ 7.23 \times 10^3~\nu^{1.1}~
{{S_C} \over {\Delta v~S_L}} {{1} \over {(1 + y^+)}} \biggr]^{0.87},$$ 

\noindent where $T^*_e$ is the LTE electron temperature
in K, $\nu$ is the observing frequency in GHz, $S_C$ is the continuum flux density
in mJy, $\Delta v$ is the half maximum full width of the line in km s$^{-1}$,
$S_L$ is the peak line flux density in mJy, and $y^+ = He^+/H^+$ is the
ionized helium to ionized hydrogen ratio, that we take equal to
0.1 (Balser et al. 1999). Using the values given in Table 1 and in the text, we
obtain an electron temperature of $T^*_e = 10,100 \pm 700$ K.
This value is in good agreement with the average values determined from optical lines.
Kaler (1986) lists nine independent determinations 
of the electron temperature
that range from 10,060 to 12,300 K, with an average value of
$11,200 \pm 600$ K, that overlaps within error our radio determination.
More recent electron temperature determinations from the Balmer
discontinuity and optical forbidden lines are also consistent with this range
(Liu \& Danziger 1993; Krabbe \& Copetti 2005; Pottasch \& Bernard-Salas 2008). 

We note that the assumption that NGC~3242 is optically thin
at 8.6 GHz is well justified because the flux density determined by us
(660 mJy) is similar to those determined by Condon \& Kaplan (1998)
at 1.4 GHz (760 mJy), by Griffith et al. (1994) at 4.9 GHz
(745 mJy), by Kaftan-Kassim (1966) at 5.0
GHz (760 mJy) and by Heckathorn (1971) at 10.6 GHz
(600 mJy), indicating that we are in the
``flat'', optically-thin part of the free-free spectrum.

\subsection{Comparison between the spatial distribution of
the continuum emission and the line emission}

\begin{figure}
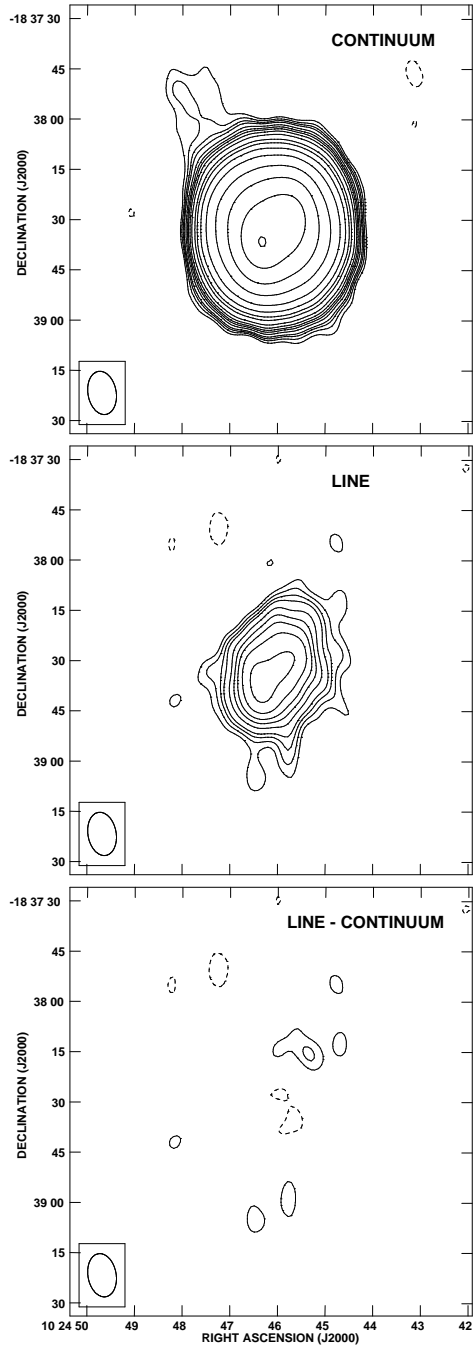

\centering
\includegraphics[scale=0.35, angle=0]{NGC3242CONT.PS}
\vskip-0.5cm
\includegraphics[scale=0.35, angle=0]{NGC3242LINE.PS}
\vskip-0.5cm
\includegraphics[scale=0.35, angle=0]{NGC3242DIFF.PS}
 \caption{Contour images of the 3.5 cm continuum
emission (top), the integrated H91$\alpha$ line emission (middle),
and the integrated line emission minus continuum difference image
(bottom), made as explained in the text. 
The contours are -3, 3, 4, 5,
6, 8, 10, 12, 15, 20, 30, 40,
60, 100, 200, 400, 800, and 1200 
times the rms noise of the images that is 0.11 mJy beam$^{-1}$ for the top image and
8.5 mJy beam$^{-1}$ km s$^{-1}$ for the middle and bottom images.} 
  \label{fig2}
\end{figure}

The good agreement of the radio-determined electron temperature with the
values from optical measurements would seem to suggest
that NGC~3242 does not have a significant continuum excess from spinning dust, 
since in this case our estimate using the electron temperature
formula given above
would have resulted in an electron temperature much larger than that
obtained in the optical measurements.

\begin{figure*}
\centering
\includegraphics[scale=0.8, angle=0]{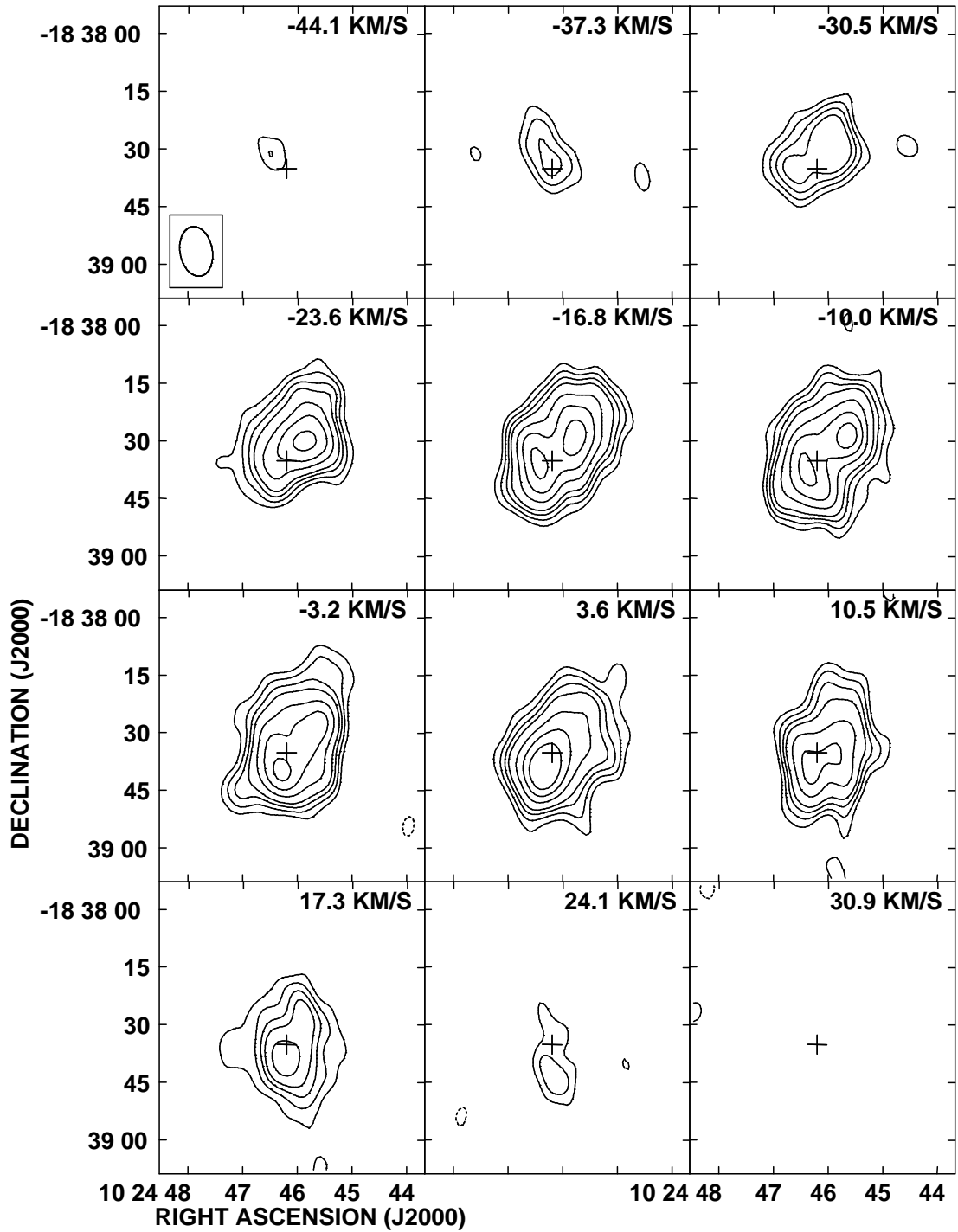}
 \caption{Channel line images of the H91$\alpha$ emission
from NGC~3242.  
The contours are -5, -4, -3, 3, 4, 5, 6, 8, 10, 12, 15, 20, and 30
times 2.04 $\mu$Jy km s$^{-1}$, the rms noise of the individual images.
The LSR radial velocity of each image is labeled in the top
right corner. The cross marks the peak position of the 
continuum emission.
The half power contour of the restoring beam ($13\rlap.{''}0 \times 8\rlap.{''}4;
PA = +10^\circ$),
is shown in the bottom left corner of the panel at $-$44.1 km s$^{-1}$.}
  \label{fig3}
\end{figure*}

There are, however, two important considerations that we discuss in what follows.
An important caveat is that the continuum excess from spinning dust
is believed to be stronger in the 10 to 30 GHz band and that it may not be
present at detectable levels at 8.6 GHz. 

A second consideration 
that can also be argued is that,
even if the spinning dust emission is present, non-LTE effects may 
enhance the radio recombination line (Dupree \& Goldberg 1970)
in such a way that this enhancement approximately compensates for the continuum increase
resulting in that the derived LTE electron temperature gives a value similar to that expected
from the optical observations.
Fortunately, we can further test this latter
possibility. In a nebula that is optically thin at centimeter wavelengths,
like NGC~3242, one expects the free-free
continuum emission and the radio recombination line emission to distribute in the sky in
a very similar manner,
since both originate basically from electron-proton interactions.
An independent
radiation mechanism such as spinning dust emission is not expected to have an
intensity distribution on the sky that follows that of the ionized gas and, if present, this 
extra component is expected to become evident in a
careful comparison between the total continuum and the radio recombination line emissions.
 
In Figure 2 we present images of the continuum and integrated line emission from NGC~3242.
The integrated line emission image was made adding the channels from $-$44.1 to +24.1
km s$^{-1}$. From a simple comparison of the images it 
is not possible to say how different are the two
images because of the much larger signal-to-noise ratio of the
continuum image, that allows to contour much more extended structure.
A first indication that the continuum and the line emissions are
very similar comes from detemining their deconvolved angular dimensions
using the task JMFIT of AIPS. For the continuum emission we
obtain $16\rlap.{''}0 \pm 0\rlap.{''}1 \times 12\rlap.{''}0 \pm 0\rlap.{''}1$;
$PA = 116^\circ \pm 1^\circ$, while for the line emission we
obtain $16\rlap.{''}1 \pm 0\rlap.{''}6 \times 11\rlap.{''}8 \pm 0\rlap.{''}1$;
$PA = 118^\circ \pm 7^\circ$. The deconvolved dimensions overlap within the error. 

To test more quantitatively
how different the two emissions are, we present in a third panel the
difference of the integrated line and
continuum images, with the continuum
image scaled (by a factor of 0.036) to have the same flux density as the integrated
line image.
As can be seen in the difference image, there is no clear signature of
a residual emission that could trace a mechanism with a different spatial
distribution to that of the free-free continuum or integrated line emissions.
The residuals of the difference image are at a $\sim$15\% level of the integrated line image
and we rule out the presence of a significant contribution from a different mechanism
at this level and at sizes 
comparable or smaller than the largest angular 
scales detectable by the VLA at 3.6 cm in the D configuration ($\sim 3'$).

Our results are consistent with those of Pazderska et al. (2009; see also
Umana et al. 2008), who failed to find excess continuum emission at 30 GHz in
their survey of planetary nebulae and concluded that free-free emission alone
can explain the observed spectra.

\begin{figure}
\centering
\includegraphics[scale=1.0, angle=0]{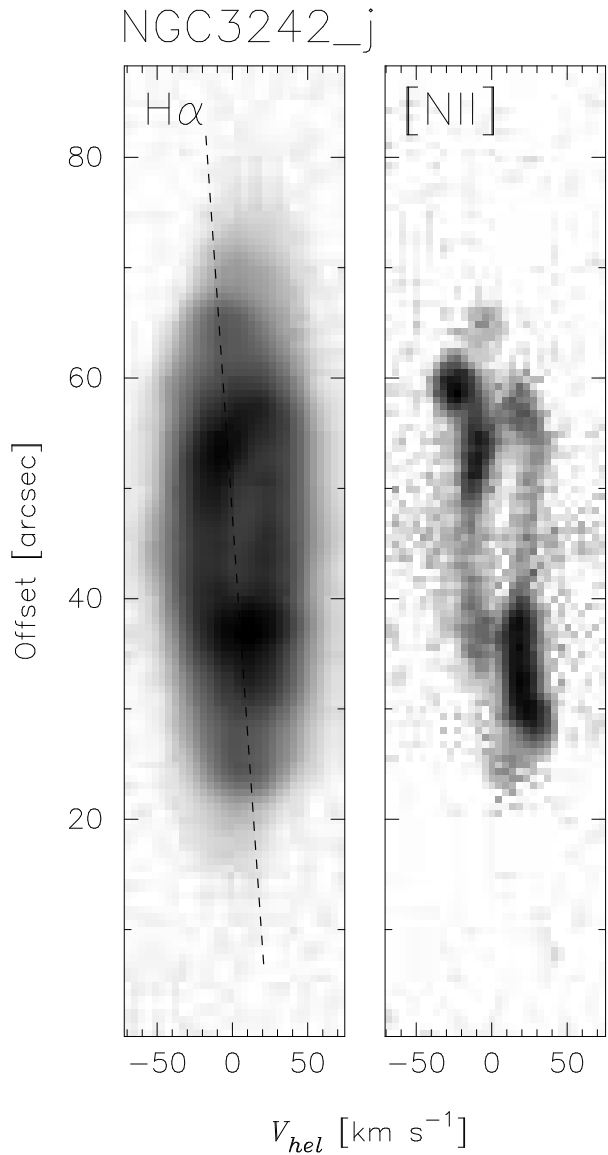}
\caption{H$\alpha$ and [N~II] $\lambda$ 6584~\AA\ bi-dimensional emission  
line spectra. The long-slit is oriented at  a  P.A. of $- 30^\circ$  
with north-west to top of the figure and south-east at the bottom. The  
dashed line in the H$\alpha$ line profile highlights the velocity  
tilt present in the bipolar outflow. The bright knots in the N~II] $\lambda$ 
6584~\AA\ correspond to the FLIERs. In the direction of NGC~3242 the heliocentric and
LSR radial velocities are related by $v_{hel} = v_{LSR} + 10.2$ km s$^{-1}$.}
  \label{fig4}
\end{figure}

\subsection{Nebular kinematics}

\subsubsection{Radio Observations}

We can also use the H91$\alpha$ observations to discuss the kinematics of the
nebula, although at the moderate angular resolution of these observations
($\sim 10{''}$).
In Figure 3 we show the individual channel line emission 
over the velocity range where there is detectable signal.
At negative LSR radial velocities, the strongest emission comes
from the NW part of the nebula, and as we progress to positive
LSR radial velocities the line emission shifts to the SE (see Fig. 3).

\subsubsection{Optical Observations}

High resolution spectroscopic observations of NGC 3242 were obtained  
at the Observatorio Astron\'omico Nacional at San Pedro M\'artir, M\'exico 
on 2009, January 15 as part of the San Pedro M\'artir  
kinematic catalog of planetary nebulae (L\'opez et al. 2006). The data  
were obtained with  the Manchester Echelle Spectrometer (MES-SPM)  
(Meaburn et al. 2003) on the 2.1 m telescope in a $f/$7.5 configuration .  
This instrument is equiped with a SITE CCD detector with 1024 $\times 
$ 1024 square pixels, each 24 $\mu$m on a side.  We used a 90 \AA{}  
bandwidth filter to isolate the 87th order containing the H$\alpha$  
and [N~II] nebular emission lines.  Two times binning was employed in  
both the spatial and spectral directions. Consequently, 512  
increments, each 0\farcs624{} long gave a projected slit length of 5\farcm32 
on the sky. We used a slit of 150 $\mu$m{} wide ($\equiv$ 11  
\kms{} and 1\farcs9) oriented along the major axis of the nebula at a  
P.A. of $-30^\circ$, and located just west of the central star. The  
exposure time was 1800 s and the spectrum was calibrated in  
wavelength against the spectrum of a Th/Ar arc lamp to an accuracy of  
$\pm$1 \kms{} when converted to radial velocity.
The bi-dimensional emission line spectra or position -- velocity (p- 
v) arrays for both H$\alpha$ and [N~II] $\lambda$ 6584~\AA\ are shown  
in Figure 4. Here the north-west corresponds to top of the profile  
and the south-east to the bottom of the figure.

As it is well known, NGC 3242 is a multi-shell elliptical nebula (Meaburn et al. 2000) with  
fast, low ionization emission knots or FLIERs (Balick et al. 1993)  
located at the leading edges of a mild bipolar outflow that emerges  
from the nucleus. The emission line profiles shown in Figure 4  
clearly reveal the main inner components of this nebula that are  
apparent in public \it{Hubble Space Telescope} \rm images. Namely, in the  
H$\alpha$ p-v array the main bright bubble is observed as the main  
velocity ellipse and the tilted bipolar outflow that is indicated in  
the figure by a dashed line,  both elements coincide with the  
kinematics mapped by our radio recombination line observations. A  
tenuous, more extended halo is also present as part of the p-v array.  
The [N~II] $\lambda$ 6584~\AA~
p-v array shows the main bright bubble and the external knots  
corresponding to the FLIERs. Clearly the tilted bipolar outflow found  
in the H$\alpha$  and the H91$\alpha$ line emissions follows the  
bipolar outflow that trails and surrounds the FLIERs.

\section{Conclusions}

We presented VLA observations of the H91$\alpha$ radio recombination
line and the adjacent continuum, as well as of H$\alpha$
and [N~II] $\lambda$ 6584~\AA, 
toward the planetary nebula NGC 3242.
Our main conclusions can be summarized as follows.

1. The
LTE electron temperature derived from the line to continuum ratio
agrees well with the values derived in the optical. 

2. The spatial distribution of continuum and line emission are very similar,
arguing against the presence of significant continuum contamination from spinning
dust emission.

3. The radial velocity structure found from the H91$\alpha$ line is consistent
with that
derived from the optical lines.


\acknowledgments
We thank S. Casassus for valuable comments. We acknowledge the support
of DGAPA, UNAM, and of CONACyT (M\'exico).
This research has made use of the SIMBAD database, 
operated at CDS, Strasbourg, France.


\end{document}